\documentclass[doublecol]{epl2}
\usepackage{amsmath,amssymb}
\usepackage{graphicx}
\newcommand{\dd}{\mathrm{d}}
\newcommand{\mitbf}[1]{\hbox{\mathversion{bold}$#1$}}

\usepackage{color}
\usepackage[normalem]{ulem} 
\renewcommand{\revision}[1]{#1}

\title{Polymer rheology simulations at the meso- and macroscopic scale }

\author{Eric Sultan$^{1,2}$, Jan-Willem van de Meent$^1$, Ell\'ak Somfai$^{1,3}$, Alexander N. Morozov$^{1,4}$ and Wim van Saarloos$^1$}

\shortauthor{E. Sultan {\em et al.}}

\institute{
\inst{1} 
{Instituut-Lorentz for Theoretical Physics, Universiteit Leiden,
P.O. Box 9506, 2300 RA Leiden, The Netherlands.}\\
\inst{2}{UPMC Univ Paris 06, Univ Paris-Sud, CNRS, UMR 7608, Lab FAST, 
Campus Univ,  F-91405, Orsay, France}\\
\inst{3}{Department of Physics and Centre for Complexity Science, University of
  Warwick, Coventry CV4 7AL, UK}\\
\inst{4} {School of Physics, JCMB, King's Buildings, University of Edinburgh, Mayfield Road, Edinburgh EH9 3JZ, Scotland}
}
\date{\today}

\abstract{We show that simulations of polymer rheology at a fluctuating mesoscopic scale and at the macroscopic scale where flow instabilities occur can be achieved at the same time with dissipative particle dynamics (DPD) technique. We model the visco-elasticity of polymer liquids by introducing a finite fraction of dumbbells in the standard DPD fluid. The stretching and tumbling statistics of these dumbbells is in agreement with what is known for isolated polymers in shear flows. At the same time, the model exhibits behaviour reminiscent of drag reduction in the turbulent state: as the polymer fraction increases, the onset of turbulence in plane Couette flow is pushed to higher Reynolds numbers. The method opens up the possibility to model nontrivial rheological conditions with ensuing coarse grained polymer statistics.}
\begin{document}
\maketitle
\section{Introduction}
The plethora of intriguing phenomena that can be observed in flows of complex fluids is attracting increasing attention among physicists. The study of polymer fluids and melts has since long occupied a central position within this field \cite{bird87v1,larson_1999}. Typically, one is either interested in the response properties of polymeric fluids or the way they flow. The former set of problems concerns viscoelasticity of polymers and biomaterials, behaviour of single long flexible molecules in flows, etcetera \cite{larson_1999,smith99,babcock_macromol}. The latter usually focuses on the differences between macroscopic flows of Newtonian and polymeric fluids in the same geometry. One such striking example is the recently discovered chaotic flows of dilute polymer solutions at very small Reynolds numbers -- the so-called {\it purely elastic turbulence} \cite{groisman2000,larson2000}. Another is the phenomenon of drag-reduction -- the observation that even minute amounts of polymer can significantly suppress Newtonian turbulence and hence reduce turbulent drag  \cite{virk}. In this Letter we develop a mesoscopic simulation method that is capable of addressing both classes of problems. 

Simulation methods of polymers essentially fall into two classes.  On one hand there are many mesoscopic coarse-grained approaches that have mainly been developed to study the thermo- and hydrodynamic properties of polymers in equilibrium and in weakly non-equilibrium situations such as an imposed small shear. Such results can, e.g., be compared with recent experimental results for the orientation statistics of single DNA molecules in solution \cite{smith99}. However, these models typically cannot be scaled up to simulate flow at macroscopic rheological scales. On the other hand, numerical studies of polymer rheology at macroscopically relevant scales are based almost exclusively on numerical implementations of continuum constitutive equations like the Oldroyd-B or FENE-P models \cite{bird87v1,larson_1999}. By their very nature, these deterministic approaches only give the average behaviour, so they cannot give insight into the coupling between the macroscopic flow behaviour and the molecular properties. In this Letter, we for the first time bridge the gap between these two approaches and scales by introducing a coarse-grained Dissipative Particle Dynamics (DPD) model \cite{hoogerbrugge_92,esp_war_95,bosch99} for visco-elastic flows which is a solution of elastic springs (dumbbells). Unlike the previous DPD simulations of polymeric fluids \cite{pan_2002}, we do not attempt to resolve internal dynamics of long polymer molecules. Instead we view the dumbbells as collective elastic degrees of freedom (normal modes) that render the solution viscoelastic. We show that the model is powerful enough to exhibit both stretching and tumbling of dumbbells at mesoscopic scales, and proper flow behaviour at hydrodynamic scales like the dramatic polymer drag reduction of turbulence.

Our approach holds an additional promise. Computational rheology still turns out to be a major challenge \cite{owens_2002}. The main difficulty lies in the fact that the convective nonlinear term in the constitutive equation can lead to a local exponential growth of the components of the polymer stress tensor \cite{hulsen}. As a result, it is difficult to come up with robust numerical schemes.  Although progress can be made by going to special variables \cite{hulsen}, there is still a great need for an easy-to-use and robust method for simulating polymer rheology in complex geometries. As is well known, mesoscale models like DPD and  Lattice Boltzmann \cite{succi}  are versatile methods to simulate laminar and turbulent Newtonian flows in rheometric flows \cite{vandemeent_2008} and in complicated geometries \cite{chen_sci_2003}; we expect our extension of the DPD model to have the same advantage.
\section{DPD model for a visco-elastic fluid}
DPD \cite{hoogerbrugge_92,esp_war_95} is an off-lattice method
in which one simulates the dynamics of particles which we can intuitively think of as representing mesoscopic blobs of fluid. The interparticle interactions consist of conserved, dissipative and random forces tuned to reproduce hydrodynamic behaviour on the scale of a few particles. Particles are assumed to have mass $m$ and velocity $\mitbf{V}_i= \dot{\mitbf{R}}_i $ at positions $\mitbf{R}_i$ so that their equations of motion are
\begin{equation}
  \label{eq:force}
  \dot{\mitbf{V}}_i =\frac{1}{m}\sum_{j\neq i} 
  \mitbf{F}^\mathrm{cons}_{ij} + \mitbf{F}^\mathrm{diss}_{ij} +
  \mitbf{F}^\mathrm{rand}_{ij} +\mitbf{F}^\mathrm{elastic}_{ij}.
\end{equation}
For the first three terms we follow \cite{vandemeent_2008} and take the standard choices \cite{hoogerbrugge_92,esp_war_95}:  the 
 force $\mitbf{F}^\mathrm{cons}$ is a soft repulsion,
\begin{equation} 
\mitbf{F}_{ij}^\mathrm{cons} = a\,\max(1-R_{ij}/R_0,0)\,\hat{\mitbf{R}}_{ij} . 
\end{equation}
Here $R_{ij}=|\mitbf{R}_i - \mitbf{R}_j|$ is the distance between the particles, $\hat{\mitbf{R}}_{ij}
$  the interparticle unit vector, and the coefficient $a$ measures the strength of this interaction. The conserved force $\mitbf{F}_{ij}^\mathrm{cons}$ increases linearly when particles come within the 
 range $R_\mathrm{0}$. From here on we take $m=R_\mathrm{0}=1$, so that 
 distances are in units of the particle radius.
The dissipative force $\mitbf{F}^\mathrm{diss}$ acts to equalise velocities of neighbouring particles, while the random force $\mitbf{F}^\mathrm{rand}$ represents a coupling to a heat bath. We use the standard form \cite{hoogerbrugge_92,esp_war_95}
\begin{align} 
  \mitbf{F}^\mathrm{diss}_{ij} &= -\Gamma
  (\hat{\mitbf{R}}_{ij}\cdot\mitbf{V}_{ij})\, \max(1-R_{ij},0)^{1/2} \hat{\mitbf{R}}_{ij} \,,\\
  \mitbf{F}^\mathrm{rand}_{ij} &= {\xi}_{ij}\sqrt{{2\Gamma
    T}/{\Delta t}} \max(1-R_{ij},0)^{1/4}\hat{\mitbf{R}}_{ij},
\end{align}
where $\Gamma$ is a friction coefficient, $T$ is the temperature, $\mitbf{V}_{ij} = \mitbf{V}_{i} - \mitbf{V}_{j}$ is the relative velocity vector, $\Delta t$ is the integration timestep, and $\xi_{ij}$ are independent Gaussian random variables with zero mean and unit variance. The form of $\mitbf{F}^\mathrm{rand}$ is chosen to ensure that the fluctuation-dissipation theorem holds in the absence of applied flow \cite{esp_war_95,chen04}. Note that the best convergence of the DPD model to hydrodynamic behaviour is achieved when all terms in Eq.(\ref{eq:force}) are roughly of the same order.

The dumbbell force $\mitbf{F}^\mathrm{elastic}_{ij}$ is our extension of  the DPD model \cite{ellak}. It is motivated by the observation that the so-called Oldroyd-B constitutive equation is exact for non-interacting dumbbells with linear elastic springs  \cite{bird87v1,larson_1999}. Linear Hookean springs, however, can lead to diverging stresses since the two partners can separate infinitely far. We therefore build our DPD model for polymer flows by using nonlinear dumbbells: we assign a unique partner to a finite fraction of the  particles and introduce, in the spirit of the Finitely Extensible Nonlinear Elastic (FENE) spring constitutive equations \cite{bird87v1,larson_1999}, a FENE-force between each particle and its partner,
\begin{equation}
\mitbf{F}_{ij}^\mathrm{FENE} = - \frac{H_\mathrm{FENE} \mitbf{R}_{ij}}{1 - (R_{ij}/R_\mathrm{max})^2}.
\label{fene_force}
\end{equation}
The FENE spring behaves as a Hookean spring with stiffness $H_\mathrm{FENE}$ at small extensions  and as a stiff rod when the extension $R_{ij}$ is close to its maximum value $R_\mathrm{max}$.

While this force works well in smooth flows, large local stresses in turbulent flows can, because of the finite timestep, sporadically lead to a blow-up due to configurations with $R_{ij} > R_\mathrm{max}$. It is therefore numerically advantageous to use instead in such situations a FENE-inspired nonlinear force that has no sharp maximum extension,
\begin{eqnarray}
\mitbf{F}_{ij}^\mathrm{SOFT}= - H_{\mathrm{{FENE}}}\,  \mitbf{R} _{ij}
\bigg[ \frac34+  \bigg(\frac14 +  \frac12\frac{R_{ij}^2}{R_{\mathrm{{c}}}^2}\bigg)
\mathrm{e}^{2{R_{ij}^2}/{R_{\mathrm{{c}}}^2}}\bigg]. \label{soft_force}
\end{eqnarray}
To the lowest order in the extension, this nonlinear force is the same as the FENE force (\ref{fene_force}) if we identify $R_\mathrm{max}$ with $R_\mathrm{c}$, but unlike Eq.(\ref{fene_force}) it remains finite for all $R_{ij}$. We will use this soft force in the last section of this Letter. Once again, we stress that the dumbbells with the force law (\ref{fene_force}) or (\ref{soft_force}) do not represent individual polymers but rather collective viscoelastic degrees of freedom.

The equations of motion are
solved with a version of the velocity-Verlet algorithm \cite{gap}. Unless noted otherwise, we take $a=1, T=0.1, \Gamma=2$, particle density $\rho=4$ and timestep $\Delta t=0.05$. Simulations are performed in the plane Couette flow geometry with periodic boundary conditions in the $x$ (streamwise) and $y$
(spanwise) direction, and no-slip walls perpendicular to the $z$
(gradient) axis.  The typical dimensions of our simulation box are $L_x{\times}L_y{\times}L_z=20{\times}20{\times}20$ so that when all particles are dumbbells  we have  $2\cdot (20)^3=16 000$ dumbbells at  density $\rho=4$. The walls are implemented as a soft repulsion
potential in the normal direction --- see \cite{ellak} for details.


\section{Rheological properties of the dumbbell DPD model} We now show that this dumbbell DPD fluid exhibits the main characteristics of polymer rheology.
In a simple shear flow, the stress tensor for an incompressible polymer solution
$\mitbf{\sigma}$ has 3 independent components: there is one independent
off-diagonal component while the diagonal components are linear combinations
of the so-called first and second normal stress differences $N_1$ and
$N_2$. While these both vanish for a Newtonian fluid, the normal
stress differences characterise the rheological
properties of a viscoelastic fluid, in particular its relaxation time
$\tau^\mathrm{rel}$. More precisely, for our plane Couette geometry we have to lowest order in $\dot{\gamma}=\partial v_x/\partial z$, 
\begin{equation}
    \sigma_{xz} = \sigma_{zx} = \eta \dot{\gamma},~~~~
  N_1\equiv |\sigma_{xx}{-}\sigma_{zz}| =2\eta_P\tau^\mathrm{rel}\dot\gamma^2, \label{rheol}
  \end{equation}
where $\eta = \eta_S + \eta_P$, with $\eta_S$ and $\eta_P$ being the solvent and the polymer contributions to the total viscosity. The expression for the first normal stress difference defines the effective polymer relaxation time  $\tau^\mathrm{rel} $. The second normal stress difference $ N_2\equiv |\sigma_{yy}{-}\sigma_{zz}| \ll N_1 $ for most polymer solutions. For the Oldroyd-B constitutive equation, $N_2$ vanishes identically \cite{bird87v1,larson_1999}.

To estimate the shear viscosity $\eta$ and the relaxation time $\tau^\mathrm{rel}$ for different dumbbell parameters, we have performed simulations at various shear rates $\dot\gamma$ and computed the components of the total stress tensor from the pairwise interactions by time-averaging the virial formula \cite{ellak,spenley00}:
\begin{align}
\mitbf{\sigma}= -\dfrac{1}{L_xL_yL_z} \sum_i\Big\{
\sum_{j\neq i}
\mitbf{F}_{ij} \otimes \mitbf{R}_{ij} 
+
\delta\mitbf{V}^z_i\otimes\delta\mitbf{V}^z_i
\Big\}
\label{virial}
\end{align}
over a time interval of length 200, which was long enough to ensure convergence; here $\delta\mitbf{V}^z_i=\mitbf{V}_i-\overline{\mitbf{V}}^z_i$ and $\overline{\mitbf{V}}^z_i$ is the average velocity of particles at the vertical position $z_i$. We have verified that only one
shear stress component takes significant values (typically
$\sigma_{xy}/\sigma_{xz}\simeq\sigma_{yz}/\sigma_{xz}\simeq0.01$), and that
$\sigma_{xx}\gg\sigma_{yy}\simeq\sigma_{zz}$ so that indeed $N_2 \ll N_1$, as desired. The accuracy of (\ref{virial}) has also been validated by comparing its results to the total momentum transferred through the walls \cite{ellak}.
\begin{figure*}
\includegraphics[width=\textwidth]{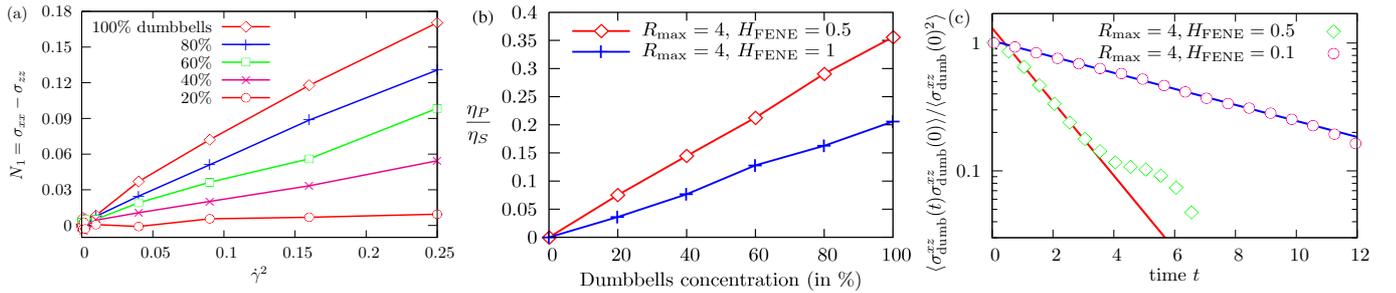}
  \caption{(colour online) (a) The normal stress difference $N_1$ as a function of $\dot{\gamma}^2$ for our DPD model (\ref{fene_force}) with $H_\mathrm{FENE}=0.5$ and $R_\mathrm{max}=4$ and various dumbbell fractions $f_P$. $\tau^\mathrm{rel}$ is determined from the fit of $N_1$ to $\dot{\gamma}^2$. (b) Viscosity ratio $\eta_P/\eta_S$ as a function of polymer fraction $f_P$ for two sets of dumbbell parameters. \revision{(c) The normalized (dumbbells contribution to the) shear stress autocorrelation function as a function of time. Lines are single-exponential fits: red -- $1.31\,\mathrm{e}^{-t/1.51}$, blue --  $1.0\,\mathrm{e}^{-t/6.1}$.}
}
    \label{N1andtau}
\end{figure*}
The viscosity $\eta$ and the relaxation time $\tau^\mathrm{rel}$ are then obtained from linear and quadratic fits of $\sigma_{xz}$ and $N_1$ to $\dot{\gamma}$ as illustrated in Fig.~\ref{N1andtau}(a). By varying the fraction $f_P$ of dumbbells from 0\% to 100\%, we can separate the individual contributions of $\eta_S$ and $\eta_P$; as shown in Fig.~\ref{N1andtau}(b), the total viscosity increases essentially linearly with the concentration, so that for the choosen DPD parameters $\eta_S\approx 0.68$ and $\eta_P \approx 0.29 f_P$.

At fixed polymer concentration, $\tau^\mathrm{rel}$ is approximately inversely proportional to $H_\mathrm{FENE}$ for large $R_\mathrm{max}$, as can be inferred from Table \ref{visc_rel}. This is as expected for the relaxation of a dumbbell in a viscous medium  \cite{bird87v1}. (We have found  that we can increase $\tau^\mathrm{rel}$  while keeping
$H_\mathrm{FENE}\simeq1$ by using chains with more than two beads
\cite{spenley00}). 
We have checked that there is little shear-thinning in our model: $\eta$
decreases by at most 10\% at high shear rates \cite{ellak}; we have also observed
that the magnitude of shear-thinning increases with $\tau^\mathrm{rel}$. \revision{We have also estimated $\tau^\mathrm{rel}$ from the decay of the shear stress autocorrelation function \cite{pan_2002}. As shown in Fig.~\ref{N1andtau}(c), it can be approximated by a single exponential that yields $\tau^\mathrm{rel}=1.51$ and $\tau^\mathrm{rel}=6.1$ for $R_\mathrm{max}=4$, and $H_\mathrm{FENE}=0.5$ and $H_\mathrm{FENE}=0.1$, respectively. These results are within 20\% of the values quoted in Table \ref{visc_rel}.}

 \begin{table}[t]
  \centering
  \begin{tabular}{||c|cc|cc|cc||}
  \hline \hline
& \multicolumn{2}{|c|}{$R_\mathrm{max}=3$}& \multicolumn{2}{|c|}{$R_\mathrm{max}=4$} & \multicolumn{2}{c||}{$R_\mathrm{max}=5$}  \\ \hline
    $H_\mathrm{FENE}$ & $\eta_P$ & $\tau^\mathrm{rel}$ 
&$\eta_P$ & $\tau^\mathrm{rel}$ &$\eta_P$ & $\tau^\mathrm{rel}$ \\ 
\hline
0.1 & 0.77 & 4.18 & 0.91  & 5.29 & 1.04 & 5.63 \\
0.125 & 0.67 & 3.07 & 0.76 & 3.55 & 0.82 & 4.22\\
0.2 & 0.49  &  2.01 & 0.55 & 2.25 & 0.58  &  2.72 \\
0.5 & 0.27 & 1.2& 0.29 & 1.23 & 0.32& 1.23 \\
1 & 0.15 & 0.64 & 0.16 & 0.66 & 0.16 & 0.65 \\ 
\hline \hline
  \end{tabular}
  \caption{Zero-shear rate fluid viscosity and relaxation time for various dumbbell parameters with $f_P=100$\% and the dumbbell FENE force (\ref{fene_force}). }
  \label{visc_rel}
\end{table}



The above results show that the dumbbell DPD model is a faithful mesoscopic representation of the Oldroyd-B type constitutive equation for polymer rheology. 

\section{Polymer stretching and orientation statistics}
The statistics of single-polymer stretching and orientation in shear flows has recently been studied both experimentally \cite{smith99} and theoretically \cite{chertkov}. We now demonstrate that our DPD dumbbell model 
is also capable of qualitatively reproducing the single-molecule results \cite{smith99,chertkov}.

\begin{figure}[h]
  \centering
  \begin{minipage}[c]{.99\linewidth}
    \centering
    \includegraphics[width=.725\textwidth]{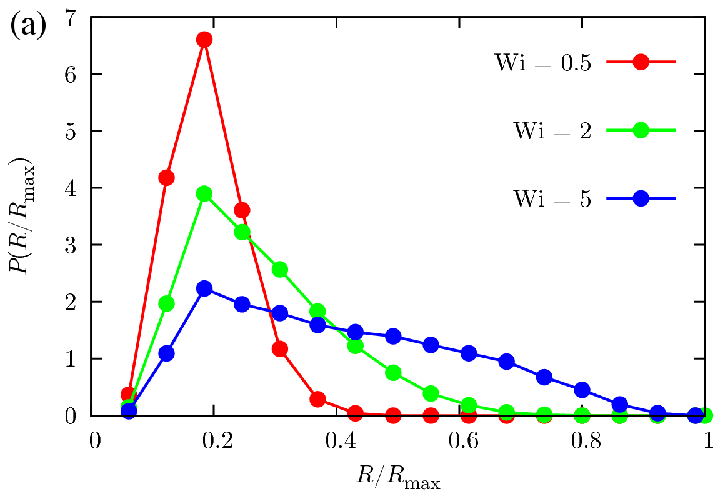}\\
          \includegraphics[width=.73\textwidth]{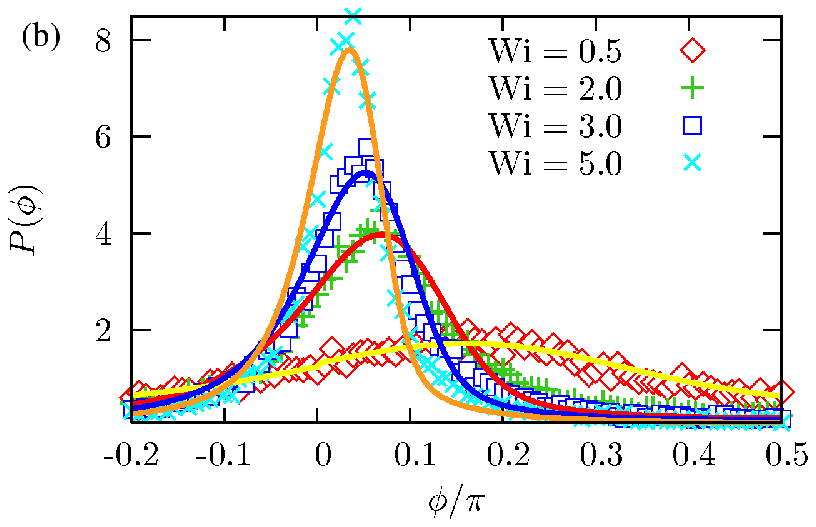}
  \end{minipage}
  \caption{(colour online) (a) Extension PDF for $\mathrm{Wi}=0.5,2$ and $5$ and $H_\mathrm{FENE}=0.5$ and $R_\mathrm{max}=5.0$. For $\mathrm{Wi}\gtrsim1$ the tail of the PDF is no longer algebraic \cite{chertkov}; this is a way to characterise the coil-stretch transition ($\mathrm{Wi}\approx 1$). (b) PDF of $\phi$ for a DPD polymer with $H_\mathrm{FENE}=1.0$ and $R_\mathrm{max}=5.0$ for different Weissenberg numbers. Continuous lines are fits to (\ref{phi_fit}). From low to high $\mathrm{Wi}$, yellow: $\dot{\gamma}/D=2.9$,  red: $\dot{\gamma}/D=41.6$, blue: $\dot{\gamma}/D=97.6$, orange: $\dot{\gamma}/D=322.9$.}
    \label{ext}
\end{figure}

 
We first study the extension of the FENE dumbbells (\ref{fene_force}) for various values of the
Weissenberg number $\mathrm{Wi}= \dot{\gamma} \tau^\mathrm{rel}$ that describes the balance between the shear flow and elasticity. Fig.~\ref{ext}(a) shows the Probability Distribution Function (PDF) of the dumbbell extension for three values of
$\mathrm{Wi}$. Upon the increase of the Weissenberg number, the shape of $P(R/R_{\mathrm{max}})$ changes very much in line with what has been found for single polymers in shear flows by other methods  \cite{celani,chertkov,babcock_macromol}: a peak at small and a power-law decay at large extensions for $\mathrm{Wi}<1$, and a rather flat distribution across a large range of extensions above the coil-stretch transition for a single polymer at $\mathrm{Wi}\approx 1$. At very high $\mathrm{Wi}$ the distribution becomes strongly peaked close to $R_\mathrm{max}$ but we have not explored this regime.

The average orientation of a polymer in a shear flow, which in our dumbbell model translates into the average orientation of the dumbbell, is characterised by the spherical angles  $\theta$ and $\phi$. The distribution of $\theta$, the angle between the dumbbell and the $xz$ shear plane, is found to be well approximated by a Lorentzian, in agreement with single-dumbbell models \cite{turitsyn05,alberto}. We focus here on the PDF of the angle $\phi$ that the projection of each dumbbell in the shear plane makes with the flow direction; in view of the symmetry of the system, we only need to consider angles $-\pi/2\leq \phi\leq\pi/2$.

The distribution $P(\phi)$ is shown in Fig.~\ref{ext}(b) together with the fits to the expression 
\begin{equation}
 P(\phi)\simeq C\int_0^{\infty}\!\mathrm{exp}\bigg[{-}\frac{\dot{\gamma}}{2D}\varphi \bigg(\frac{1}3\varphi^2-\frac{\sin2\phi}2\varphi+\sin^2\phi\bigg)\bigg]\dd\varphi ,
\label{phi_fit}
\end{equation}
which can be derived from the Fokker-Planck equation for $P(\phi,t)$ \cite{turitsyn05}. In this expression $D$ plays the role of a diffusion coefficient for the orientation angle $\phi$. As can be seen from the figure, the above expression fits our data very well, and this allows us to study the shear-rate dependence of $D$; from the fitted values of $\dot{\gamma}/D$ listed in the caption, one infers that $D$ decreases with increasing shear rate approximately as   $\dot{\gamma}^{-1.3}$. Although for a FENE model one does not expect a pure scaling form, we can compare this to what is expected for a single dumbbell with a linear spring, for which  $D\sim\dot\gamma^{-2}$, as well as to what is expected for inextensible rods,  for which $D = const$ \cite{celani}. This comparison indicates that over the range of shear rates considered, the finite extensibility of the dumbbells already reduces the shear rate dependence of $D$ noticeably.

\section{Tumbling statistics} As we noted before, the orientation statistics of an isolated polymer can largely be understood by studying a dumbbell in a shear flow \cite{chertkov}.  This also holds for the tumbling statistics  \cite{smith99,briels2005}: in a shear flow, the polymer is most of the time almost aligned with the mean flow, but due to the torque exerted on it by shear, every now and then it rapidly tumbles before coming back close to the average orientation; the tumbling time $\tau^\mathrm{tum}$ is defined as the typical time between two such events \cite{smith99,briels2005}. In our DPD model we have many dumbbells which interact via $F^\mathrm{cons}$. Nevertheless, we shall now show that our model still exhibits the same tumbling statistics as a single dumbbell in shear flow. 

\begin{figure}[t]
  \centering
  \begin{minipage}[c]{.999\linewidth}
    \centering
    \includegraphics[width=.75\textwidth]{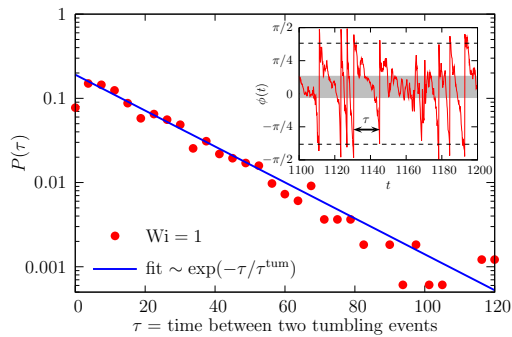}
  \end{minipage}
\caption{Distribution of tumbling times for 6 polymers in shear flow at $\mathrm{Wi}=1$ during 10000 simulation time units (providing typically 2000 tumbling events). Dumbbell parameters are $H_\mathrm{FENE}=1$ and $R_\mathrm{max}=4$. The distribution is well fitted by an exponential. \emph{Inset:} Corresponding time trace of the azimuthal angle of a dumbbell. We take the tumbling time $\tau$ to be the time between two large values of $\phi$ (outside the dashed blue lines $|\phi|=1.2$) such that $\phi$ takes at least one value close to  the average (inside the grey region)  for some intermediate time.}
    \label{tumblage}
\end{figure}

To compare the tumbling time $\tau^\mathrm{tum}$ with the relaxation time
$\tau^\mathrm{rel}$, we performed DPD simulations in a $18{\times}18{\times}18$
box and other parameters as before and followed the dynamics of $6$ dumbbells for $10000$ time units. We computed their extension $R (t)$ and azimuthal angle $\phi(t)$  every $3\Delta t$. The intermittent nature of such time traces is immediately clear from the example shown in the inset of Fig.~\ref{tumblage}.

In Table \ref{table_rel} we list the average tumbling times $\tau^\mathrm{tum}$ for different values of $R_\mathrm{max}$ and $H_\mathrm{FENE}$. The values of $\tau^\mathrm{tum}$ were computed from the angular dynamics as explained in the caption of Fig.~\ref{tumblage}. We have also determined the individual tumbling times from the time interval between two stretched configurations with a contracted one in between. Both methods yield comparable results, but contrary to single polymer experiments for which angular resolution is a limitation \cite{gerash_prl}, analysing the angular time traces works best numerically and is least dependent on the thresholds used to define a tumbling event.

\begin{table}[t]
\begin{center}
\begin{tabular}{||l|cc|cc||}
\hline \hline
 &\multicolumn{2}{|c|}{$R_\mathrm{max}=4$}&\multicolumn{2}{c||}{$R_\mathrm{max}=5$}\\ \hline
$H_\mathrm{FENE}$&  $0.5$ & $1$& $0.5$ & $1$\\
$\tau^\mathrm{tum}$& 29.80 & 20.36  & 29.20 & 23.01 \\
\hline \hline
\end{tabular}
\end{center}
\caption{Single dumbbell tumbling times at $\mathrm{Wi}=1$ for various dumbbells stiffness and maximal extension. Essentially the same values are found for all Weissenberg numbers up to $\mathrm{Wi} \approx 5$, except that we find a slight decrease with increasing shear rate for the parameters corresponding to the rightmost column. }
\label{table_rel}
\end{table}

In our simulations, the distribution of tumbling times shown in Fig.~\ref{tumblage} decays exponentially --- this is known to be a robust property of a single polymer in shear flow \cite{celani,alberto}. The fact that there is only one intrinsic relaxation time in our model is another important illustration that its viscoelastic behaviour is a good mesoscopic representation of the single-relaxation time in Oldroyd-B type constitutive equations. Qualitatively, the collective tumbling of DPD polymers is  very much like what one expects from Brownian dynamics of a single dumbbell: the tumbling time increases with the fluid relaxation time (we almost have proportionality for $R_\mathrm{max}=5$). As expected \cite{celani}, we find that $\tau^\mathrm{tum}\propto\tau^\mathrm{rel}$ for not too large $\mathrm{Wi}$.


\begin{figure}[t]
  \centering
\includegraphics[width=.85\linewidth]{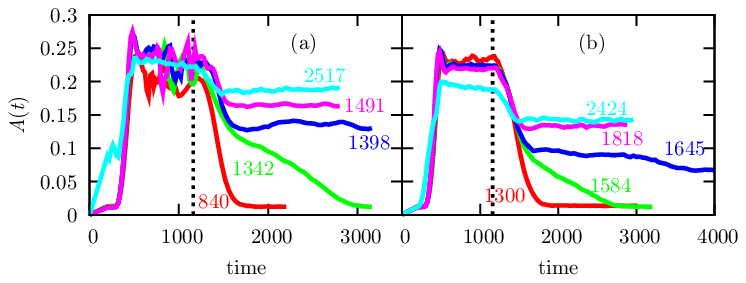}\\
         \includegraphics[width=.775\linewidth]{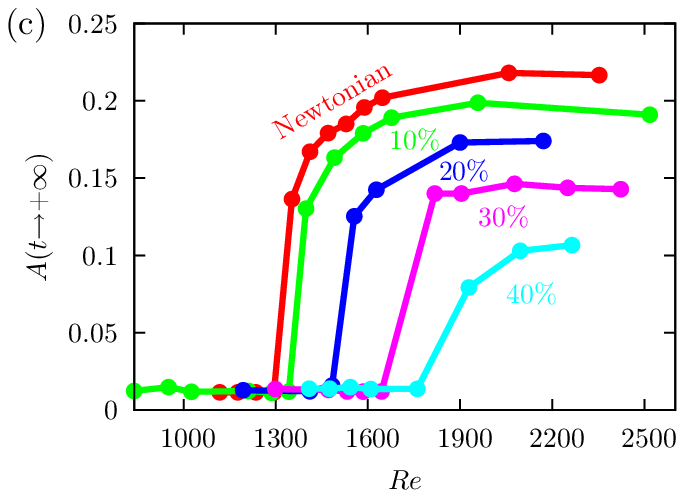}
         \vspace*{-3mm}
         
\caption{(a,b) Plots of the turbulent amplitude $A(t)$ versus time as a function of Reynolds number ${Re}$ for $10\%$ (left) and $30\%$ (right) polymer fractions. The Reynolds number $Re$ is given besides each curve. The dotted line corresponds to the time at which we start switching off the driving force. (c) Influence of polymers on the stability of turbulent state in our dumbbell DPD simulations. As the fraction of polymers is increased, the onset of turbulence is shifted to higher values of $Re$ and the turbulent amplitude is reduced.} 
    \label{Aversust}
\end{figure}

We have also performed simulations with a very few dumbbells -- only 12 out of the  $8\cdot (18)^3$ particles were paired up with FENE springs. Except for small string stiffness $(H_\mathrm{FENE}\leq0.1)$ (presumably because of the poor statistics and the force singularity at large stretching), calculation of the associated tumbling times gave values very close to those with $100\%$ dumbbells listed in Table \ref{table_rel}. Together with the results of the previous section, this shows that the elasticity of the surrounding dumbbells does not significantly affect local dynamics. It implies that by increasing the concentration of dumbbell polymers $f_P$, we can increase the total normal stress without changing the properties of the fluid on a mesoscopic scale. 

\section{Turbulent drag reduction}
To demonstrate that our approach can also be employed to study non-trivial macroscopic flows, we perform, for the first time to our knowledge, DPD simulations of high Reynolds number turbulent flows in the presence of polymers. This system is known to exhibit drag-reduction -- the phenomenon that the addition of minute amounts of polymer significantly suppresses turbulence and hence the drag \cite{virk}. It has recently been successfully captured by direct numerical simulations \cite{suresh97,dubief2004} and by studying the exact coherent states \cite{stone2004}.
We have shown previously \cite{vandemeent_2008} that the standard DPD model reproduces the main characteristics of turbulent plane Couette flow (albeit the onset Reynolds numbers are somewhat too high perhaps due to the significant compressibility of the DPD fluid). We now study the effect of polymers on the turbulent state as a function of polymer fraction. 

To do so, we now switch to the soft FENE dumbbell interaction (\ref{soft_force}) for the reasons we explained before. The parameters for these runs are $a=4$, $\rho=4$, $T=0.1$, $\Gamma=1$, $H_\mathrm{FENE}=0.5$ and $R_\mathrm{c}=2$ (the rheological properties are $\eta_S\approx0.69$, $\eta_P\approx0.25f_P$ and $\tau_\mathrm{rel}\approx0.9$); the simulation box has dimensions $L_x\times L_y\times L_z=60\times40\times20$\footnote{In these runs, the walls are modelled by a layer of continuous DPD matter (see \cite{vandemeent_2008}), rather than by a smooth potential.}.

\begin{figure}
  \centering
     \includegraphics[width=.95\linewidth]{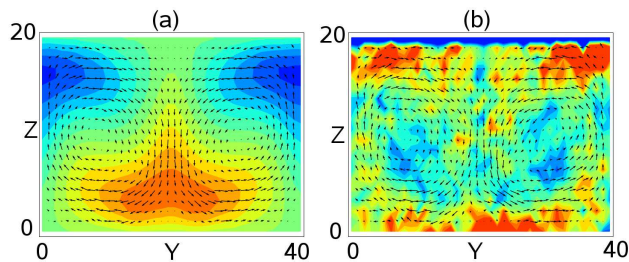}
\vspace{-3mm}

  \caption{(colour online) Streamwise velocity and polymer extension in the gradient-spanwise plane in a simulation with $f_P=10\%$ and $Re=2520$. Colours represent the streamwise deviation (locally averaged) from (a) the laminar velocity  (vectors denote the in-plane motion) and (b) the average dumbbell stretching. Colour code is such that red and blue indicate a $\pm 79\%$ deviation from the average velocity, and for stretching a variation of $\pm 66\%$ about the average stretch ($\approx 1.1$).}
\label{stretch}
\end{figure}

We first drive the flow as in \cite{vandemeent_2008} with a force field that generates an array of streamwise vortices that are known to dominate the turbulent state at least close to the onset \cite{eckhardt_arfm_2007}. The driving field is turned on at time 200, then slowly ramped up to time 440, kept at a steady value till time 1160, and then slowly turned off till time 1400. By following the maximum deviation $A(t)$ from the linear flow field, $A(t)=\mbox{max}(\bar{V} (z)-\dot{\gamma}z)$ where $\bar{V}(z)$ is the spatially averaged mean streamwise velocity profile, it is then tested whether turbulence remains  or whether it decays to the laminar state. The behaviour of $A(t)$ was studied as a function of polymer fraction $f_P$ and the Reynolds number ${Re}$. We have varied the forcing amplitude over a factor 12 and found the final turbulent state (if it exists) to be independent of the forcing amplitude. Typical data for two polymer concentrations is shown in Figs.~\ref{Aversust}(ab). The saturated values of $A(t$$\to$$\infty)$ are plotted in Fig.~\ref{Aversust}(c) as a function of ${Re}$ for various $f_P$. This stability diagram illustrates the occurrence of drag reduction within our dumbbell DPD model: for increasing polymer fraction, the onset of turbulence is pushed to higher Reynolds numbers and the turbulent amplitude is reduced. A similar stability diagram was reported for the exact coherent states in plane Couette flow \cite{stone2004}.
Finally, as Fig.~\ref{stretch} illustrates, not only do the dumbbells get preferentially stretched inside the streaks (blue and red regions in the velocity plot), as one might expect \cite{dubief2004}, but the fluctuations in the stretching appear to be much larger than those in the velocity. 


\section{Conclusion} 
In this paper we have shown that the dumbbell DPD model has all the main characteristics of the Oldroyd-B class constitutive equations, while the statistical properties of the dumbbells are similar to those of single polymers. It can thus be used to study polymer rheology simultaneously at the macroscopic and mesoscopic scales. This makes the DPD dumbbell model an attractive candidate for simulations of polymer rheology in complex geometries and flows, in particular in situations where impact of the macroscopic flow on the mesoscopic statistical properties is of interest. We expect the model to be especially useful and easy to implement for studies of cross-slot geometries, flows past objects, extrusion, etc. Detailed information about coarse grained polymer conformation readily available in our model may even shed new light on the mechanism of drag reduction.

\acknowledgments
The authors acknowledge NWO/NCF for SARA supercomputer time (Project No. MP-06-119). The work of ES and ANM in Leiden was supported by the physics foundation FOM (ES) and the DoP program of NWO/FOM (ANM). ANM acknowledges support from the Royal Society of Edinburgh/BP Trust Personal Research Fellowship.

\bibliography{polymers}
\end{document}